\documentclass[10pt,journal]{IEEEtran}
\usepackage[font=small,labelsep=space]{caption}
\usepackage{verbatim}
\usepackage{graphicx}
\usepackage{subfig}
\usepackage{multirow}
\usepackage{amssymb}
\usepackage{amsmath}
\usepackage{amsthm}
\usepackage{amsfonts}
\usepackage{paralist}
\usepackage{url}
\usepackage{epstopdf}
\usepackage{float}
\usepackage[justification=centering]{caption}
\usepackage{color}
\usepackage{paralist}
\usepackage[noend]{algpseudocode}
\usepackage{algorithmicx}
\usepackage{algorithm}

\let\itemize\compactitem
\let\enditemize\endcompactitem
\let\enumerate\compactenum
\let\endenumerate\endcompactenum

\begin{document}
\title{Human-Perception-Oriented Pseudo Analog Video Transmissions with Deep Learning}

\author{Xiao-Wei~Tang,~\IEEEmembership{Student Member,~IEEE}, Xin-Lin~Huang*,~\IEEEmembership{Senior Member,~IEEE}, \\
Fei Hu,~\IEEEmembership{Member,~IEEE}, and Qingjiang Shi

\thanks{Xiao-Wei Tang (email: {\tt xwtang@tongji.edu.cn}) is with the Department of Control Science and Engineering, Tongji University, Shanghai 201804, China.}
\thanks{Xin-Lin Huang (email: {\tt xlhuang@tongji.edu.cn}) is with the Department of Information and Communication Engineering, Tongji University, Shanghai 201804, China. (corresponding author)}
\thanks{Fei Hu (e-mail: {\tt fei@eng.ua.edu}) is with the Department of Electrical and Computer Engineering, University of Alabama, Tuscaloosa, AL 35487 USA.}
\thanks{Qingjiang Shi(e-mail: {\tt shiqj@tongji.edu.cn}) is with the School of Software Engineering, Tongji University, Shanghai 201804, China.}
}

\maketitle

\begin{abstract}
Recently, pseudo analog transmission has gained increasing attentions due to its ability to alleviate the cliff effect in video multicast scenarios. The existing pseudo analog systems are optimized under the minimum mean squared error criterion. However, their power allocation strategies do not take the perceptual video quality into consideration. In this paper, we propose a human-perception-based pseudo analog video transmission system named ROIC-Cast, which aims to intelligently enhance the transmission quality of the region-of-interest (ROI) parts. Firstly, the classic deep learning based saliency detection algorithm is adopted to decompose the continuous video sequences into ROI and non-ROI blocks. Secondly, an effective compression method is used to reduce the data amount of side information generated by the ROI extraction module. Then, the power allocation scheme is formulated as a convex problem, and the optimal transmission power for both ROI and non-ROI blocks is derived in a closed form. Finally, the simulations are conducted to validate the proposed system by comparing with a few of existing systems, e.g., KMV-Cast, SoftCast, and DAC-RAN. The proposed ROIC-Cast can achieve over 4.1dB peak signal-to-noise ratio gains of ROI compared with other systems, given the channel signal-to-noise ratio as -5dB, 0dB, 5dB, and 10dB, respectively. This significant performance improvement is due to the automatic ROI extraction, high-efficiency data compression as well as adaptive power allocation.
\end{abstract}

\IEEEpeerreviewmaketitle

\section{Introduction}
\IEEEPARstart {W}{ith} the development of mobile terminals, a large number of wireless video applications have emerged, such as augmented reality/virtual reality (AR/VR), unmanned aerial vehicle (UAV) video surveillance, 4K live video streaming, and so on [1]. In this paper, we mainly focus on the video multicast scenario characterized by low delay and high quality. The two biggest challenges facing the video multicast scenario are the rigorous delay requirement and the drastically fluctuating channel conditions [2].

The conventional digital systems typically divide a video clip into groups of pictures (GoPs) and compress the GoPs into bit stream through a standard video encoder (e.g., JPEG 2000, H.264/AVC, etc.) [3]. Intra- and inter-frame correlations of the video enable high compression efficiency. Therefore, the amount of data transmitted is greatly decreased, thus reducing the end-to-end transmission delay. To deal with the drastically fluctuating channel conditions, the conventional digital systems adopt an adjustable channel modulation/coding scheme (MCS) to overcome the channel interference based on the channel state feedback [4]. However, the selected MCS may not guarantee a predetermined packet loss ratio (PLR) due to the fluctuations of the channel conditions. The cliff effect will occur under deep channel fading conditions [5]. Especially in the multicast scenario, the transmitter needs to select a MCS according to the worst channel quality of receivers to guarantee the correct demodulation in all the receivers [6]. Thus, today's wireless digital  systems are ineffective in video multicast scenarios.

Recently, pseudo analog transmission technology has attracted many attentions due to its ability to alleviate the cliff effect in video multicast scenarios [7, 8]. In [7], it has been proved that the analog system could achieve the same channel capacity as the digital system under the minimum mean squared error (MMSE) criterion. In [8], the first pseudo analog video transmission scheme named SoftCast was proposed, in which the transmitter does not need to know the receiver's channel capacity/quality since the demodulation quality is proportional to the users' corresponding channel quality. With the current advancement in 5G and the abundance in data rate, Liu et al. proposed to achieve high video recovery performance by leveraging the MIMO-OFDM technology and pseudo analog transmission technology [9].

However, the existing pseudo analog transmission systems are optimized solely based on the MMSE criterion without taking the perceptual video quality into consideration. According to the research on human visual perception, people can only pay attention to one or a few areas at a time when watching videos, and the area of concentration is called region of interest (ROI) [10]. The existing pseudo analog systems typically allocate transmission power to DCT coefficients according to their variances. However, specific content-focused missions are needed to first guarantee the perceptual quality of ROIs. For example, in the island patrol missions, the details of invading ships are more eye-catching than the sea level since those details can help the police quickly detect an anomaly; in traffic video surveillance missions, the moving vehicles and pedestrians are more noticeable than the background such as roads, trees, and buildings; in equipment monitoring missions, people are more interested in the color of the indicator light which represents different operation states. Therefore, in this paper we will investigate how to enhance the transmission quality of ROI part in content-focused missions through the pseudo analog transmission technology.

How to accurately extract the ROI part from the transmitted image/video is the primary problem to be solved in this research. Inspired by the development of \emph{saliency detection} in video compression, we propose to apply saliency detection in pseudo analog video transmissions to extract the salient objects for subsequent ROI coding [11]. Conventional saliency detection methods often adopt hand-crafted low-level features to differentiate the foreground and background with some specific separation models [12-14], including principal component analysis (PCA) based low-rank decomposition model [12], Gaussian mixture model [13], and other color/texture-based models [14]. Generally, these methods have low accuracy and can only work on specific data sets. Recently, deep learning (DL) has been used for saliency detection due to its fast processing speed, high  detection accuracy as well as strong adaptability [15-17]. In [15], a fully convolutional neural network called DeepFix was proposed to automatically learn the features in a hierarchical fashion and predict the location of the saliency objects in an end-to-end manner. In [16], Szegedy et al. investigated various pre-trained models in terms of feature extraction for saliency detection, including AlexNet, VGG-16, and GoogLeNet. In [17], a regression-based saliency detection method named YOLOv2 was proposed, which used the whole image as the input of the network and generated the positions of the salient objects.

In this paper, we will propose a novel {\textbf{ROI-C}}haracterized pseudo analog multi{\textbf{cast} }system named ROIC-Cast to enhance the transmission quality of ROI part in content-focused missions. In the proposed ROIC-Cast, the video frames will be firstly decomposed into ROI and non-ROI blocks based on the DL-based saliency detection model. The non-ROI blocks will be treated as the background, while ROI blocks are delivered via the routing paths with better channel quality and higher transmission power. To our best knowledge, this is the first work that seamlessly integrates DL algorithm into the pseudo analog video transmission. To sum up, the main difference between our work and the existing work is that we optimize the content-focused system by taking the perceptual video quality of ROI part into consideration, not just based on the MMSE criterion. \\
\textbf{Contributions}: Our contributions can be concluded as the following three-fold:
\begin{enumerate}
\item The classic YOLOv2 algorithm is adopted in the ROI extraction to automatically decompose the video sequences into ROI and non-ROI blocks at the transmitter.
\item An effective compression method is proposed to reduce the data amount of side information generated by ROI detection.
\item the power allocation scheme is formulated as a convex problem where the optimal transmission power for both ROI and non-ROI blocks is given in a closed form.
\end{enumerate}

The rest of the paper is organized as follows. Section II provides a brief review of related work including YOLOv2 and pseudo analog transmission technology. In Section III, three main components of the proposed ROIC-Cast framework are described in details, including ROI extraction, side information compression, and unequal power allocation. The implementation details and simulation results, as well as the comparisons with conventional schemes are covered in Section IV, followed by the conclusions in Section V. All the variables/notations to be used in the remaining sections are listed in Table I.

\begin{table}[htbp!]
\centering
\caption{NOTATION LIST}
\begin{tabular*}{8.5cm}{c|l}
\hline
Notation &Meaning\\
\hline
$S$ &the length of grid cells in YOLOv2 \\
$(x,y)$ & the coordinate of the center of the bounding box\\
$(w,h)$ & the width and height of the bounding box\\
{\emph{confidence}} & the credibility of the detected object \\
$\mathcal B$ &the number of bounding boxes and confidences   \\
$B_i$ & the $i^{th}$ DCT coefficient block \\
$g_i$ & the power scaling factor of $B_i$\\
$\lambda_i$ & the variance of $B_i$\\
$P_t$ & the total transmission power \\
$K_i$ & the correlation factor between $B_i$ and the similar block\\
$\ell ({K_i})$ & the correlation coefficient\\
$M$ & the size of the $B_i$ \\
$\mathcal {F}$ & the whole input image \\
$t_i$ & the $i^{th}$ detected salient object\\
$B_s$ & the first block of the detected object\\
$B_e$ & the last block of the detected object\\
$H$  &the height of the transmitted frame   \\
$W$  &the width of the transmitted frame  \\
$P_s$ & the transmission power allocated for side information\\
$\beta$ & the corresponding SNR that meets \\
&the lowest BLER requirement\\
$\sigma_0^{ - 2}$  &the variance of the channel noise  \\
$\theta_i$ & the transmitted signal set of $B_i$ \\
$N_s$ & the total number of blocks in a frame\\
$r_i$ & the received signal set of $B_i$\\
$n$ & the additive white Gaussian noise\\
$D_i$ & the reconstruction distortion of $B_i$\\
$P_d$ & the transmission power allocated for DCT coefficients\\
$P_{dr}$ &the transmission power allocated \\
&for ROI DCT coefficients\\
$P_{dnr}$ &the transmission power allocated \\
&for non-ROI DCT coefficients\\
$S(r)$  & the size of ROI blocks\\
$S(nr)$ & the size of non-ROI blocks\\
$\eta$ & the ratio between the average transmission power \\
&for ROI coefficients and non-ROI coefficients\\
$N_r$ & the number of ROI blocks\\
$N_nr$ & the number of non-ROI blocks\\
\hline
\end{tabular*}
\label{T1}
\end{table}

\vspace{-1mm}
\section{Related Work}
In this section, we will give a brief review of YOLOv2 and pseudo analog transmission technology. Firstly, we will briefly summarize the superiorities of YOLO series over other DL-based saliency detection algorithms, and highlight the differences among YOLO series. Then, we will make a literature review of several well-known pseudo analog schemes, and introduce the power optimization strategies in these pseudo analog schemes.

\vspace{-2mm}
\subsection{YOLOv2 Algorithm}
YOLO series (including YOLOv1 [18], YOLOv2 [17], and YOLOv3 [19]) are the state-of-the-art saliency detection algorithms  proposed by Redmon J \emph{et al}. in [17-19]. YOLO series treat saliency detection as a regression problem, in which the whole image is regarded as the input of the network. The positions, categories, and corresponding confidence probabilities of all salient objects contained in the input can be obtained through an inference process in YOLO series.

YOLO series have many superiorities over other DL-based saliency detection algorithms, e.g., R-CNN [20], Fast R-CNN [21], and Faster R-CNN [22]. Firstly, the training and detection processes in YOLO series are implemented together in an end-to-end neural network whereas R-CNN series adopt two separated steps to obtain the candidate boxes. Therefore, the YOLO series are faster and more accurate than R-CNN series. Specifically, YOLO series are capable of processing 67 pictures per second with $76.8\%$ detection accuracy. Secondly, YOLO series have lower background error detection rate compared to R-CNN series. Thirdly, YOLO series have stronger versatility. YOLO series' detection rate of non-natural image objects is much higher than R-CNN series.

The performance difference within YOLO series is also obvious in terms of detection speed and detection rate. For example, YOLOv1 has higher detection accuracy for large objects than small ones. Compared to YOLOv1, YOLOv2 can run at various image sizes, offering an easy tradeoff between speed and accuracy. YOLOv3 addresses the shortcomings of YOLOv1 and YOLOv2 which have low detection accuracy for small objects. However, the network structure of YOLOv3 is more complicated than YOLOv2, thus the detection speed is not as fast as YOLOv2. To achieve real-time wireless video communications, we choose YOLOv2 algorithm to perform the saliency detection in this paper, whose neural network structure is shown in Fig. 1.

\begin{figure*}[htbp!]
\centering
\includegraphics[width=1\textwidth]{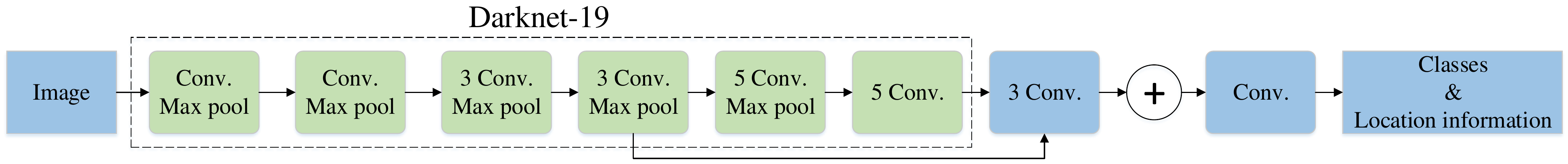}
\caption{The architecture of YOLOv2.}
\label{F1}
\end{figure*}

The YOLOv2 network is an extended version of Darknet-19 classification network structure, which is composed of 22 convolutional layers and 5 pooling layers (see Fig. 1). Unlike R-CNN series, the output layer of YOLOv2 is no longer a {\emph{softmax}} function, but a {\emph{tensor}}. Specifically, YOLOv2 divides the input image into $S \times S$ grid cells in the training process and each grid cell predicts whether or not the center of an object falls into its interior. If a grid cell predicts that the center of an object falls within it, the grid cell continues to predict $\mathcal B$ bounding boxes and $\mathcal B$ confidences for the object. Each bounding box contains five parameters: $x$, $y$, $w$, $h$, and {\emph{confidence}}. Specifically, $(x,y)$ represents the coordinate of the center of the bounding box, $(w,h)$ represents the width and height of the bounding box, and {\emph{confidence}} reflects the credibility of the object detected by the bounding box.

\vspace{-2mm}
\subsection{Pseudo Analog Transmission}
Many novel pseudo analog transmission schemes have been proposed in recent years. In [8], Katabi et al. proposed a cross-layer design for wireless video broadcast named SoftCast, which was the first work on pseudo analog transmission. SoftCast removes quantization and entropy coding from the conventional digital systems, and directly transmits the power-scaled discrete cosine transform (DCT) coefficients. Specifically, DCT coefficients with larger variances are allocated with more transmission power. In addition, when the bandwidth is not sufficient, SoftCast could discard the least important DCT coefficients (i.e., those with smaller variances) to save bandwidth.

In [23], it indicated that the correlation of videos could be fully utilized via a pseudo analog scheme called D-Cast. In D-Cast, the received frames are regarded as the\emph{ side information} to assist with the reconstruction of current frames. This greatly improves the energy efficiency. A data-assisted cloud radio access network for visual communications, named DAC-RAN, was proposed in [24]. It separates the control and data planes in the conventional digital transmission infrastructure, and integrates a new data plane (that is specifically designed for video communications) into the virtual base station. The correlated information retrieved from video big data is utilized as the prior knowledge in video reconstruction. However, the quality of the reconstructed video does not increase linearly with the increase of signal-to-noise ratio (SNR), due to mutual interference. In [25], He et al. proposed a structure-preserving video delivery system named SharpCast to improve both the objective and subjective visual quality. In [26], Huang et al. proposed a knowledge-enhanced wireless video transmission system called KMV-Cast which could exploit the hierarchical Bayesian model to integrate the correlated information into the video reconstruction. In [27], a \emph{maximum a posteriori} (MAP) decoding was proposed for KMV-Cast pseudo analog video transmission to further eliminate the residual noise in the received video/image. In [28], the well-known block-matching and three dimensional filtering (BM3D) algorithm were adopted to remove the noise for KMV-Cast. The above studies [25-27] mainly focused on reducing the effect of noise on the demodulation quality of image/video at the receiver, while ignoring the effect of video content on quality of experience (QoE) performance.

Next, we use SoftCast and KMV-Cast as the examples to introduce the power optimization concept in pseudo analog transmission schemes. In SoftCast, all DCT coefficients are divided evenly into blocks (i.e., $B_i$) with a uniform size (e.g., $8 \times 8$). Each DCT coefficient block $B_i$ is assigned with a power scaling factor ${g_i}$ according to its variance in order to obtain the symbols satisfying the power constraints. In SoftCast, the optimal power scaling factor ${g_i}$ can be denoted as [8]:
\begin{equation}\label{E6}
{g_i} = {\lambda _i}^{ - 1/4}\left( {\sqrt {\frac{{{P_t}}}{{\sum\nolimits_i {\sqrt {{\lambda_i}} } }}} } \right),
\end{equation}
where ${P_t}$ denotes the total transmission power and $\lambda_i$ denotes the variance of $B_i$. From Eqn. (1), one can conclude that the transmission power for $B_i$ is determined by its variance in SoftCast.

There are two main differences between KMV-Cast and SoftCast: 1) In order to reduce the transmission delay, each block is de-correlated through 2D-DCT transform instead of 3D-DCT transform; 2) KMV-Cast can make full use of the video resources already stored in the cloud to assist with the video demodulation at the receiver. Specifically, assume that the transmitted images/videos share the same statistical distribution with the reference ones stored in the cloud database. It has been proved in [26] that the performance of KMV-Cast is affected by the correlation factor $K_i$ which represents the similarity between the transmitted block $B_i$ and the reference block. Similar to SoftCast, the optimal power scaling factor in KMV-Cast can be denoted as [26]:
\begin{equation}\label{E10}
{g_i} = {\lambda _i}^{ - 1/4}\left( {\sqrt {\frac{{{P_t}\sqrt {\ell({K_i})} }}{{\sum\nolimits_i {\sqrt {{\lambda_i}} \sqrt {\ell({K_i})} } }}} } \right),
\end{equation}
where $\ell ({K_i})$ represents the correlation coefficient which is only related to $K_i$ and can be denoted as [26]:
\begin{equation}\label{E9}
\ell ({K_i}) = \left({K_i} + \sqrt {(M - 1)(1 - {K_i^2})}\right)^2,
\end{equation}
where $M$ represents the size of the block $B_i$ which is a constant.

Please note that the side information (i.e., $K_i$ and $\lambda_i$) is transmitted in digital mode due to their indispensability to the video reconstruction at the receiver, whereas the scaled DCT coefficients are transmitted in pseudo analog mode. Comparing to Eqn. (2), one can find that the scaling factor of KMV-Cast is determined by both $\lambda_i$ and $K_i$.

\section{Framework of ROIC-Cast.}
In this section, we will describe the details of three main steps of the proposed ROIC-Cast system. Those three steps are: ROI extraction, side information compression, and unequal power allocation. The system model is shown in Fig. 2. Firstly, ROI is extracted from each frame with YOLOv2 structure. Then, each video frame is divided into blocks uniformly, and each block is de-correlated through 2D-DCT transform. Next, the blocks are categorized into ROI and non-ROI ones, and an unequal power allocation scheme is proposed to enhance the transmission quality of the ROI blocks. Finally, the DCT coefficients and the compressed side information are mapped into the I/Q components of the transmitted signals.
\begin{figure*}[htbp!]
\centering
\includegraphics[width=1\textwidth]{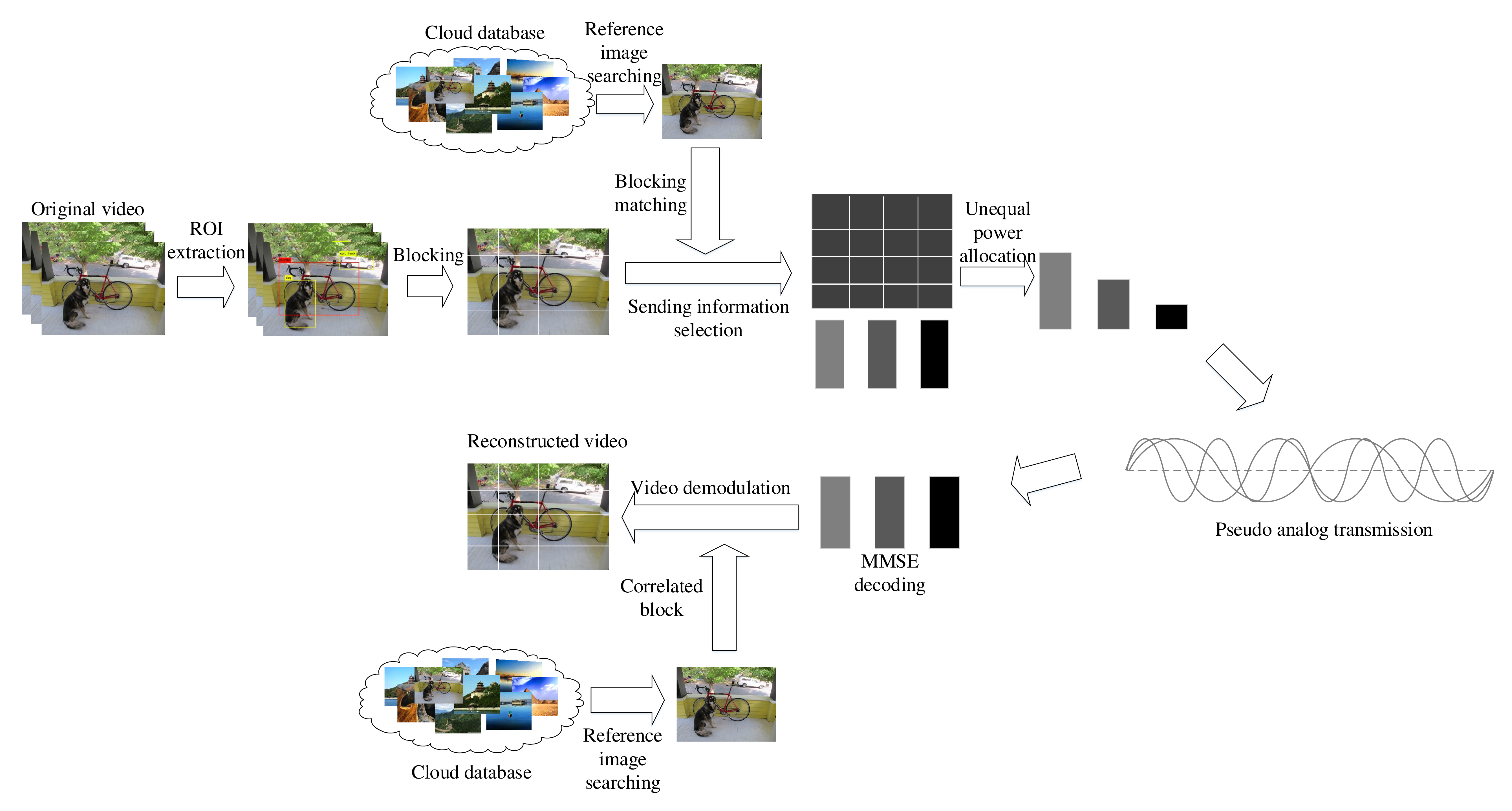}
\caption{Overview of the proposed ROIC-Cast scheme.}
\label{F2}
\end{figure*}

\vspace{-2mm}
\subsection{ROI Extraction}
Let $\mathcal {F} :\mathcal {F}  \to \mathbb{R}$ be a whole frame and ${\mathcal {F}^{ROI}} \subset \mathcal {F}$ be the ROI part contained in frame $\mathcal {F}$. In ${\mathcal {F} ^{ROI}}$, we usually have a set of salient objects with the shape of rectangles, and each object $t_i$ has a center $(x_i,y_i)$, width $w_i$ and height $h_i$. At the output layer of YOLOv2, the position information (i.e., $x_i,y_i,w_i,h_i$) of each object $t_i$ can be obtained. These position information can help us effectively locate the ROI at the receiver, as shown in Fig. 3(a). Note that each video frame is divided into blocks at uniform size (e.g., $8\times8$).
\begin{figure}[htbp!]
\centering
\includegraphics[width=0.5\textwidth]{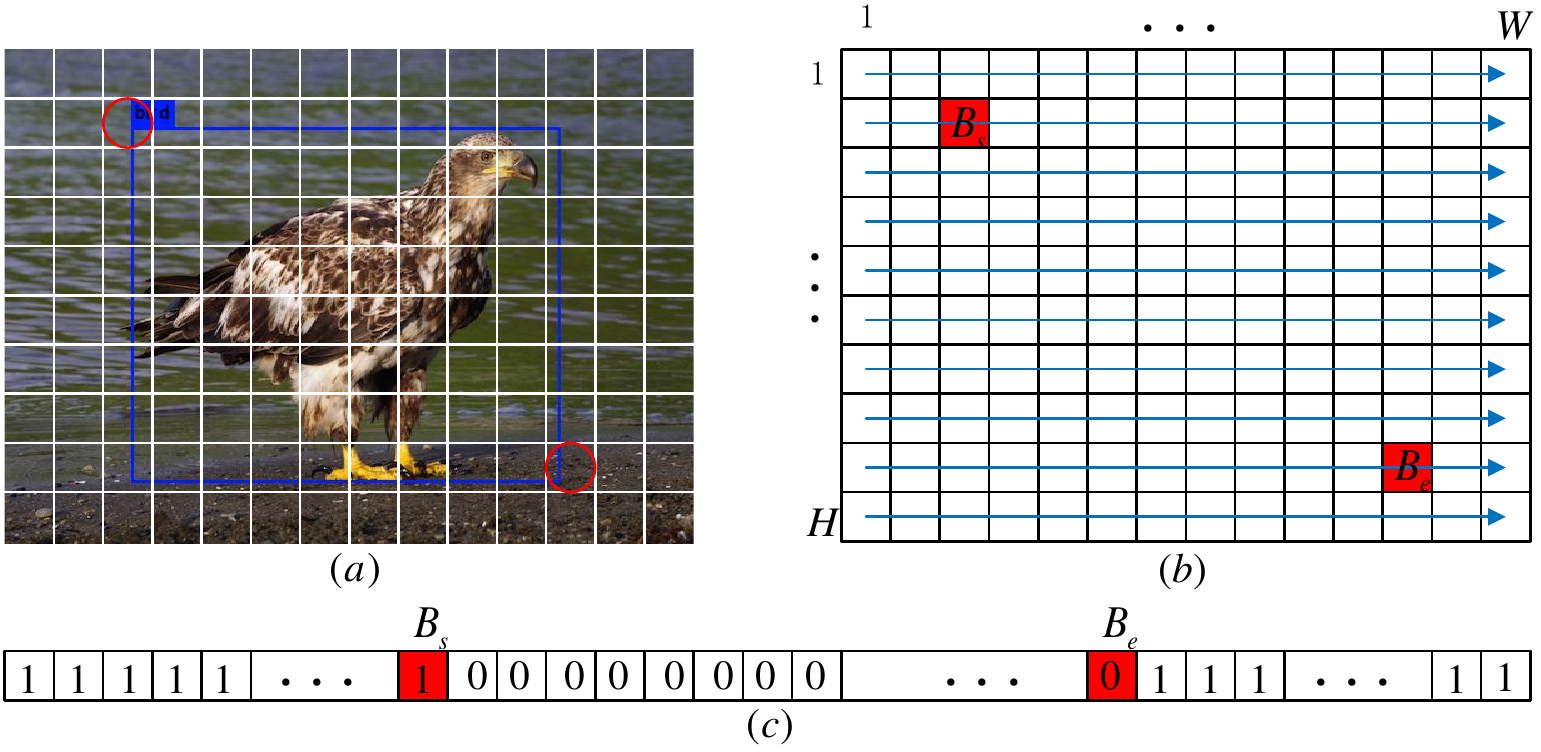}
\caption{Run-length coding of ROI position information.}
\label{F3}
\end{figure}

In order to transmit the location information concisely, we further decrease the data amount and compress the position information of ROI by using run-length coding (RLC). Specifically, RLC is used to record the positions of the starting and ending blocks of each object $t_i$. We assume that the size of the video frame is $H \times W$ ($H$ and $W$ represents the height and the width of the video frame, respectively). From Fig. 3(b), one can see that the ROI can be determined by just the sequential numbers of the blocks $B_s$ at the left upper corner and $B_e$ at the right lower corner of the ROI block. That is, the ROI location can be represented as Fig. 3(c). The first red bit $1$ represents the starting block $B_s$ and the second red bit $0$ represents the ending block $B_e$.

\vspace{-2mm}
\subsection{Side Information Compression}
In the pseudo analog video multicast system, we have pointed out that the side information ($K_i$ and $\lambda_i$) is needed at the receiver to assist with video reconstruction (see Section II, Part B).

To prevent the side information from bit errors, the side information is coded in a digital mode and the standard LTE system is used to transmit the side information. The LTE system can adjust the MCS adaptively, according to the block error ratio (BLER) desired by the system. In [29], 15 different channel quality indicators (CQI) are defined, where different CQI values correspond to different MCSs. The smaller the CQI value is, the stronger the protection for the transmitted information will be, but the number of bits that can be carried by each symbol will also decrease. In the case of insufficient bandwidth, it is necessary to discard some high-frequency DCT coefficients with small variances to meet the bandwidth requirements when using the MCS with low CQI.

In ROIC-Cast, there are three kinds of side information. The first one is the ROI location information (i.e., $B_{s}$ and $B_{e}$), which can tell the receiver where the ROI is. The second one is the variance of the pixel block (i.e., $\lambda_i$), which has a large impact on the visual quality of the reconstructed video. The third one is the correlation factor (i.e., $K_i$) between the transmitted block and the reference block stored in the cloud. All the side information (i.e., $B_{s}$, $B_{e}$, $\lambda_i$, and $K_i$) is first compressed into bitstream using Huffman coding (note that FPGA-based Huffman coding can be extremely efficient, and its processing time is at the nanosecond level.) [30]. Since Huffman coding can only use integers to represent a single symbol, there will be irreversible quantization errors in the practical transmission process.

The pseudo analog transmission is often used in multicasting and fast fluctuation channels, and the system must protect the side information which should be decoded as correctly as possible. In order to serve most users, we choose the MCS which can satisfy the lowest BLER requirement. Assume that $\sigma_0^2$ represents the variance of the channel noise, the transmission power allocated for the side information must satisfy:
\begin{equation}\label{E11}
{P_s} \triangleq \beta {\sigma_0^2},
\end{equation}
where $\beta$ represents the corresponding SNR that meets the lowest BLER requirement. Eqn. (4) shows the required minimum power (i.e., ${P_s}$) for the side information in ROIC-Cast transmission system.

\vspace{-2mm}
\subsection{Unequal Power Allocation}
The ROIC-Cast scheme proposed in this paper can be applied in the scenarios with relatively static background, such as video surveillance applications which mainly detect moving objects in a static background. Therefore, the background of the whole video remains almost the same. Long-term repetitive reconnaissance can be used to build a comprehensive cloud database [31,32], which can provide reference blocks while transmitting the videos. Thus the transmission quality of non-ROI blocks can be guaranteed by fully utilizing the reference blocks provided by the cloud database and more transmission power can be used to protect the transmission quality of ROI blocks. Therefore, the power allocation scheme of ROIC-Cast needs to make a trade-off between ROI and non-ROI blocks, which is more challenging than that of KMV-Cast.

Similar to the aforementioned pseudo analog systems, we assume that all DCT coefficients in ROIC-Cast are divided evenly into blocks with $M$ DCT coefficients in each block. The DCT coefficients in $B_i$ can be regarded as independent samples generated by a zero-mean Gaussian variable, which can be denoted as $\{\theta_i|{\theta_i}\sim {\mathcal{N}}(0,{\lambda _i}),i=1,2...,N_s\}$, where $N_s$ represents the total number of blocks in a frame and $\lambda_i$ represents the variance of $B_i$. Each $B_i$ is assigned with a power scaling factor ${g_i}$ in order to obtain the symbols satisfying the power constraints. For ease of derivation, we assume that the signals are transmitted in an additive white Gaussian noise (AWGN) channel. Then, the received signal (i.e., ${r_i}[m], m = 1, ...,M$) can be expressed as:
\begin{equation}\label{E2}
{r_i}[m] = {g_i} \cdot {\theta_i}[m] + n,
\end{equation}
where ${\theta_i}[m]$ represents the  $m^{th}$ coefficient in $B_i$, and $n$ represents the AWGN with variance of $\sigma_0^2$. Thus, the decoded signal can be denoted as:
\begin{equation}\label{E3}
{\hat \theta_i}[m] = \frac{{{r_i}[m]}}{{{g_i}}} = {\theta_i}[m] + \frac{n}{{{g_i}}}.
\end{equation}

Then the expected reconstruction distortion of each block $B_i$ can be denoted as:
\begin{equation}\label{E4}
\begin{array}{l}
D_i = E\left[ {\sum\limits_{m=1}^{M} {{{({\hat \theta_i}[m] - {\theta_i}[m])}^2}}} \right],\\
\;\;\;\;\; = {M} {\frac{{E[{n^2}]}}{{g_i^2}}},\\
\;\;\;\;\; = {\frac{{{M} \sigma_0^2}}{{g_i^2}}}.
\end{array}
\end{equation}

Since ROI blocks are more important than non-ROI blocks, the quality degradation of ROI blocks has a big impact on people's understanding of the video contents. Thus, more power should be allocated for the ROI blocks. Assume that the transmission power for each frame is a constant denoted as ${P_{t}}$. Then, the transmission power for the DCT coefficients (i.e., $P_d$) can be denoted as:
\begin{equation}\label{E12}
{P_{d}} = {P_{t}} - {P_{s}},
\end{equation}
where $P_s$ represents the power allocated for the side information whose expression has been given in Eqn. (4).

Let $P_{dr}$ denote the transmission power allocated for ROI DCT coefficients, and $P_{dnr}$ denote the transmission power for non-ROI DCT coefficients. Thus, ${P_d}$ can be also denoted as:
\begin{equation}\label{E13}
{P_d} = {P_{dr}} + {P_{dnr}}.
\end{equation}

In this paper, we define a preference parameter $\eta$ ranging from 0 to 1, which represents the ratio between 1) the average transmission power for each non-ROI pixel and 2) the average transmission power for each ROI pixel. The definition of $\eta$ is:
\begin{equation}\label{E14}
\eta  = {{\frac{{{P_{dnr}}}}{{S(nr)}}} \mathord{\left/
 {\vphantom {{\frac{{{P_{dnr}}}}{{S(nr)}}} {\frac{{{P_{dr}}}}{{S(r)}}}}} \right.
 \kern-\nulldelimiterspace} {\frac{{{P_{dr}}}}{{S(r)}}}} = \frac{{{P_{dnr}}}}{{{P_{dr}}}} \cdot \frac{{S(r)}}{{S(nr)}},
\end{equation}
where $S(r)$ represents the size of ROI blocks and $S(nr)$ represents the size of non-ROI blocks. Both $S(r)$ and $S(nr)$ can be calculated at the receiver according to the well-protected side information. From Eqn. (10), one can see that the transmission power allocated for ROI DCT coefficients increases with the increase of $\eta$.

According to Eqn. (9) and Eqn. (10), $P_{dr}$ and $P_{dnr}$ can be derived as:
\begin{equation}\label{E15}
{P_{dr}}  = \frac{{S(r)}}{{\eta S(nr){\text{ + }}S(r)}} \cdot {P_d}. \\
\end{equation}
\begin{equation}\label{E15}
{P_{dnr}} = \frac{{\eta S(nr)}}{{\eta S(nr){\text{ + }}S(r)}} \cdot {P_d}. \\
\end{equation}

In ROIC-Cast, the goal is to minimize the distortion of both ROI and non-ROI of the received video frames. Thus, its optimization functions are similar to that in KMV-Cast which can be formulated as follows:
\begin{equation}\label{E17}
\left\{ \begin{gathered}
  \mathop {\min }\limits_{{g_i^2}} \sum\limits_{i = 1}^{{N_r}} {\ell ({K_i})D_i},{\rm{s.t.}}\left\{ \begin{gathered}
  \sum\limits_{i = 1}^{{N_r}} {{g_i^2}{\lambda _i}\ell ({K_i})}\leqslant{P_{dr}} \hfill \\
  {g_i^2}{\lambda _i}\ell ({K_i})\geqslant 0 \hfill \\
\end{gathered}  \right.,if {B_i}\!\in\!\mathcal{F}^{ROI}, \hfill \\
  \mathop {\min }\limits_{{g_i^2}} \sum\limits_{i = 1}^{{N_{nr}}} {\ell ({K_i})D_i},{\rm{s.t.}}\left\{ \begin{gathered}
  \sum\limits_{i = 1}^{{N_{nr}}} {{g_i^2}{\lambda _i}\ell ({K_i})} \leqslant {P_{dnr}} \hfill \\
  {g_i^2}{\lambda _i}\ell ({K_i})\geqslant 0 \hfill \\
\end{gathered}  \right.,otherwise. \hfill \\
\end{gathered}  \right.
\end{equation}
where $N_{r}$ represents the number of ROI pixel blocks and $N_{nr}$ represents the number of non-ROI pixel blocks. $\ell({K_i})$ is only related to $K_i$, which is already expressed in Eqn. (3).

Note that the above-mentioned two parallel sub-problems are both convex optimization problems regarding to the variable $g_i^2$. Therefore, we can apply Karush-Kuhn-Tucker (KKT) conditions for the optimal solution [33]. Specifically, we can derive the optimal power scaling factor for ROIC-Cast in a closed form according to the Lagrange multiplication method as follows [34]:
\begin{equation}\label{E18}
{g_i} = \left\{ \begin{gathered}
  {\lambda _i}^{ - 1/4}{\sqrt {\frac{{{P_{dr}}\sqrt {\ell ({K_i})} }}{{\sum\limits_{i = 1}^{{N_r}} {(\sqrt {{\lambda _i}} \sqrt {\ell ({K_i})} )} }}} },\;\;if\;{B_i} \in {\mathcal {F}^{ROI}},\hfill \\
  {\lambda _i}^{ - 1/4}{\sqrt {\frac{{{P_{dnr}}\sqrt {\ell ({K_i})} }}{{\sum\limits_{i = 1}^{{N_{nr}}} {(\sqrt {{\lambda _i}} \sqrt {\ell ({K_i})} )} }}} },\;otherwise. \hfill \\
\end{gathered}  \right.
\end{equation}

From Eqn. (14), one can see that the optimal power scaling factor of block $B_i$ are determined by three aspects, including 1) the variance $\lambda_i$, 2) the correlation factor $K_i$, and 3) whether or not the block $B_i$ belongs to the ROI. Compared with the power allocation strategies of SoftCast and KMV-Cast (as shown in Eqn. (1) and Eqn. (2)), one can conclude that the proposed unequal power allocation can provide more protection for ROI coefficients that have large effect on QoE performance. The proposed scheme might have a slightly higher computation overhead than some of the existing works due to the extraction of the ROI part from the video via YOLOv2 algorithm. However, with today's high-performance computing hardware/software, such an additional computation overhead should be a minor issue.

The detailed solution process of the proposed unequal power allocation scheme is provided in Algorithm 1.
\begin{algorithm}[t]
\caption{Proposed unequal power allocation algorithm.}
\hspace*{0.02in} {\bf Input:} Reference frame, transmission frame, preference parameter $\eta$, total transmission power $P_{t}$, and noise power variance $\sigma _0^{ - 2}$.\\
\hspace*{0.02in} {\bf Output:} Optimal power allocation $g_i$.
\begin{algorithmic}[1]
\State \textbf{ROI extraction}: Extract the ROI from the transmitted frame using YOLOv2, and record the position information (i.e., $B_s$ and $B_e$) of the detected salient objects.
\State \textbf{Source processing}: Divide the transmitted frame into blocks at uniform size (e.g., $8\times8$), de-correlate each block through 2D-DCT transform, and calculate the variance of each block (i.e., $\lambda_i$).
\State \textbf{Block matching}: Find the reference block that is most similar to the transmitted block, and calculate their correlated factor (i.e., $K_i$).
\State \textbf{MCS selection}: Compress the side information (i.e., $B_s$, $B_e$, $\lambda_i$, and $K_i$) using Huffman coding, select the MCS which satisfies the predetermined BLER requirement, and calculate ${P_{s}}$ according to Eqn.(4).
\State \textbf{Power allocation}: Calculate ${P_{d}}$, ${P_{dr}}$, and ${P_{dnr}}$ according to Eqn. (8), Eqn. (11) and Eqn. (12), respectively, and then calculate ${g_i}$ according to Eqn.(14).
\State \Return ${g_i}$
\end{algorithmic}
\end{algorithm}

\vspace{-2mm}
\section{Performance Analysis}
We first provide the implementation details for the standard LTE transmission system. Then the parameter settings are given. Finally we analyze the simulation results and compare our scheme with other three typical pseudo analog schemes.
\vspace{-2mm}
\subsection{Implementation Details}
\begin{figure*}[htbp!]
\centering
\includegraphics[width=1\textwidth]{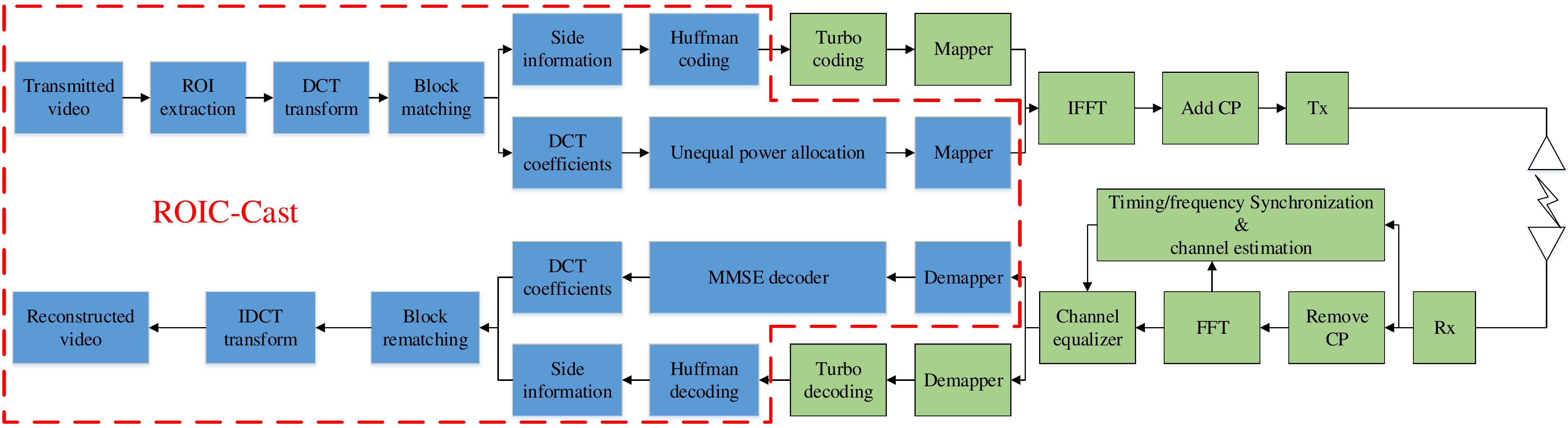}
\caption{The implementation scheme based on the LTE system.}
\label{F4}
\end{figure*}
In the standard LTE system (with the bandwidth of $1.4$MHz), the channel is divided into 64 subcarriers, of which 4 subcarriers are used to transmit pilot signals for channel estimation, and 48 subcarriers are used to transmit user data. In ROIC-Cast, we separate the transmitted video into two parts including the side information and the DCT coefficients. The side information is first compressed into bit stream by using Huffman coding. Then, according to the required BLER, the corresponding MCS is selected to modulate the side information into complex signals. Afterwards, we map each group of two DCT coefficients to I/Q to generate a complex signal. IFFT transform is then performed for those complex signals generated by the side information and the DCT coefficients together. At the receiver, the DCT coefficients and side information are restored after signal synchronization, channel estimation, channel equalization, and FFT transform. Finally, the video is reconstructed with the side information and DCT coefficients. The specific implementation details are shown in Fig. 4, where the modules of ROIC-Cast have been marked with a red, dashed box.

\vspace{-2mm}
\subsection{Parameters settings}
We have used three natural video sequences (e.g., ``Coastguard'', ``Highway'', and ``Container'') and three non-natural video sequences (e.g., ``Foreman'', ``Akiyo'', and ``Hall'') to verify the effectiveness of the proposed algorithm. These video sequences are open source and have been widely used for simulation analysis in multimedia research [35-41]. The details of these six video sequences are provided in Table II, including the resolution, frame number, frame rate, and size. All of them have a 8-level pixel depth (i.e., the pixel value ranges from 0 to 255).

In standard video sequences, there is usually a high spatial-temporal correlation between frames. The closer the video frames are, the higher the correlation is, and \emph{vice versa}. Therefore, we take the first frame of each video sequence as the reference which can provide similar blocks, and we take the second frame and the $180^{th}$ frame to simulate the strong correlation case and the weak correlation case, respectively. Both the reference frame and the transmitted frames will be evenly divided into uniform blocks with $M=64$ DCT coefficients in each blocks.
\begin{table}[htbp!]
\centering
\caption{Details of testing video sequences.}
\begin{tabular}{ccccc}
\hline
Sequences  &Resolution &Frame number & Frame rate& Size \\
\hline
Coastguard &$176 \times 144$ &300& 25fps& 5.74MB \\
\hline
Highway &$176 \times 144$ &2000& 25fps& 36.3MB \\
\hline
Container &$176 \times 144$ &300& 25fps& 4.06MB \\
\hline
Foreman &$176 \times 144$ &300& 25fps& 5.88MB \\
\hline
Akiyo &$176 \times 144$ &300& 25fps& 1.86MB \\
\hline
Hall &$176 \times 144$ &300& 25fps& 5.72MB \\
\hline
\end{tabular}
\label{T2}
\end{table}

In the simulations, the BLER is set to $10^{-3}$ [42]. We test the performance of all the schemes over AWGN channel with $\sigma _0^{ - 2} =10^{-3}$, and set the channel's SNR range from $-5$dB to 25dB. Table III shows the corresponding MCSs that satisfy the BLER requirement under different channel conditions in LTE system, where ECR stands for an effective code rate. In this paper, we ensure that all the schemes share the same bandwidth (i.e., the number of the complex signals to be transmitted is the same) and the same transmission power. When the bandwidth is insufficient, the high-frequency components of DCT coefficients are discarded to meet the bandwidth budget.
\begin{table}[htbp!]
\centering
\caption{The MCSs adopted for different channel conditions.}
\begin{tabular*}{5.5cm}{lclcl}
\hline
$\ \ \beta$ & ~~CQI& Modulation & ECR \\
\hline
-5dB &\ \ ~1 &\ \ 4QAM &\ \ 0.0762  &\ \  \\
\hline
 ~0dB&\ \ ~4 &\ \ 4QAM &\ \ 0.3008  &\ \  \\
\hline
 ~5dB &\ \ \ 7 &\ \ 16QAM &\ \ 0.3691  &\ \   \\
\hline
10dB  &\ \ \ 9 &\ \ 16QAM &\ \ 0.6016 &\ \  \\
\hline
15dB  &\ \ \ 12 &\ \ 64QAM&\ \ 0.6504  &\ \  \\
\hline
20dB  &\ \ \ 15 &\ \ 64QAM &\ \ 0.9258  &\ \  \\
\hline
\end{tabular*}
\label{T2}
\end{table}

We analyze the performance of different schemes in terms of the peak signal-to-noise ratio (PSNR) [43] and the subjective visual quality. The PSNR is a standard objective measurement of video/image quality, and is defined as a function of the $M\!S\!E$ between all pixels of the reconstructed video and the original version as follows [43]:
\begin{equation}\label{E1}
P\!S\!N\!R =  20{\log}(\frac{{255}}{{\sqrt {M\!S\!E} }}),
\end{equation}
where $M\!S\!E$ represents the mean square error of the reconstructed image.

\vspace{-2mm}
\subsection{Performance Analysis of ROI Extraction}
The ROI extraction results determine the performance of the proposed ROIC-Cast scheme. In this part, we investigate the visual quality of $12$ reconstructed video frames, and evaluate the performance of ROI extraction via YOLOv2 structure{\footnote{YOLOv2 structure can be referred at http://pjreddie.com/darknet/yolov2.}}. The results are shown in Fig. 5.
\begin{figure*}[htbp!]
\centering
\includegraphics[width=1\textwidth]{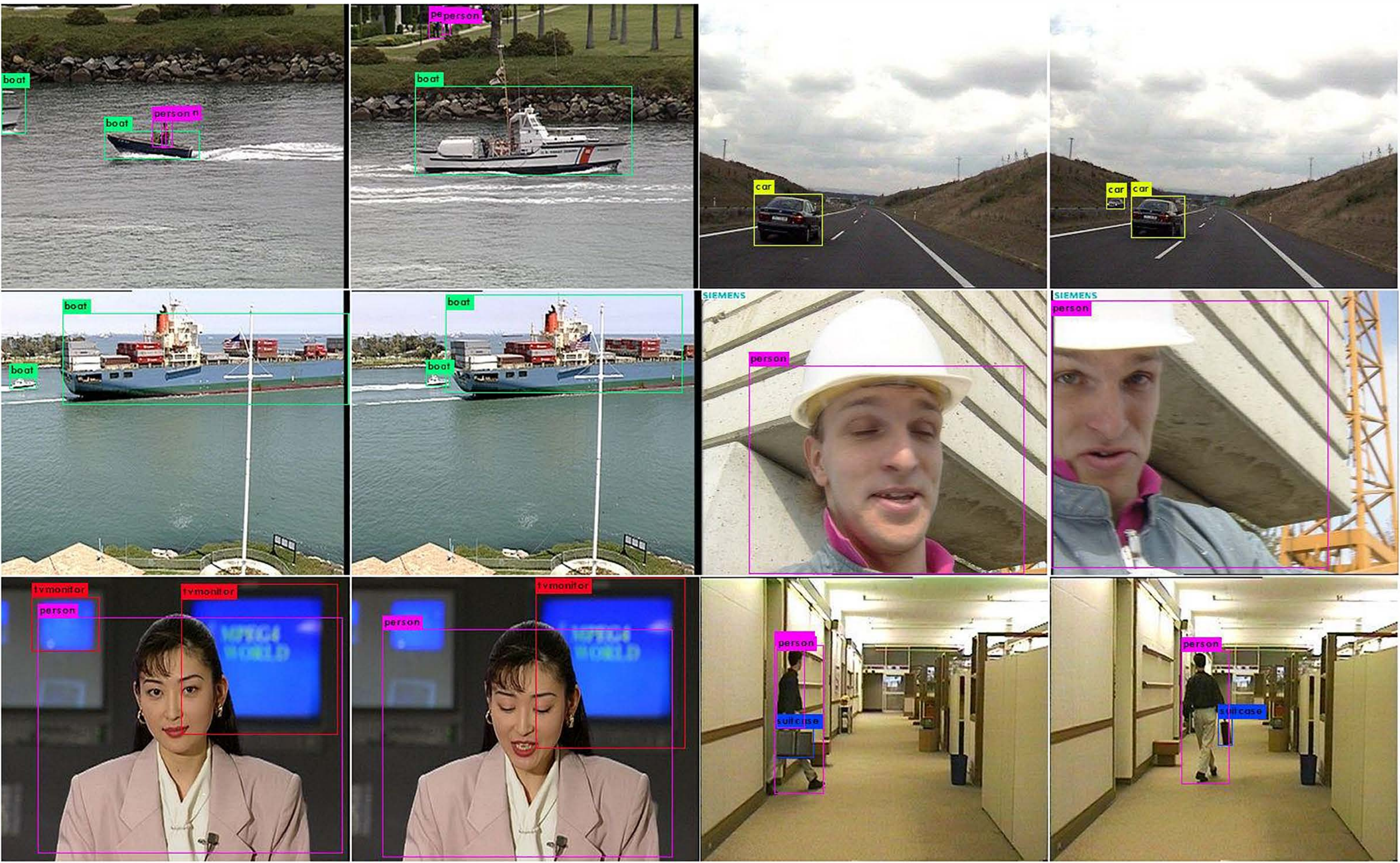}
\caption{ROI extraction results via YOLOv2 structure. The first row: (a) Coastguard \#2; (b) Coastguard \#180; (c) Highway \#2; (d) Highway \#180; The second row:(a) Container \#2; (b) Container \#180; (c) Foreman \#2; (d) Foreman \#180; The third row: (a) Akiyo \#2; (b) Akiyo \#180; (c) Hall \#2; (d) Hall \#180.}
\label{F5}
\end{figure*}

In Fig. 5, the three rows show the ROI extraction results of 12 selected frames via YOLOv2 structure. In Coastguard \#2, the ROI consists of three salient objects, e.g., a complete boat, a person, and a part of another boat (note that those two boats are independent of each other). In Coastguard \#180, besides the clearly visible boat and the person on it, YOLOv2 can even identify the person walking on the shore. Two cars are detected which are driving in opposite directions on different roads in Highway \#180 while only one car appears in Highway \#2. In Container \#2 and \#180, two boats including a yacht and a freighter are detected. The face of a man is recognized as the ROI in Foreman \#2 and \#180. In Akiyo \#2 and \#180, there are mainly three salient objects, i.e., two TV monitors and one person, which overlap with each other. A man with a suitcase appears in Hall \#2 and \#180, whose posture and position are changing over time. From Fig. 5, one can conclude that YOLOv2 can extract the ROIs from each video frame accurately, which is effective for both natural and non-natural videos.

\vspace{-2mm}
\subsection{Performance Analysis of Unequal Power Allocation}
To investigate the performance of the proposed unequal power allocation algorithm for ROI and non-ROI data, we observe the overall PSNR of the reconstructed images as well as the PSNR of the ROI, under different values of the preference parameter $\eta$ (see Eqn. (10)). The channel SNR is set to -5dB, 0dB, 5dB, and 10dB, respectively. The results are shown in Fig. 6.
\begin{figure}[htbp!]
\centering
\includegraphics[width=0.4\textwidth]{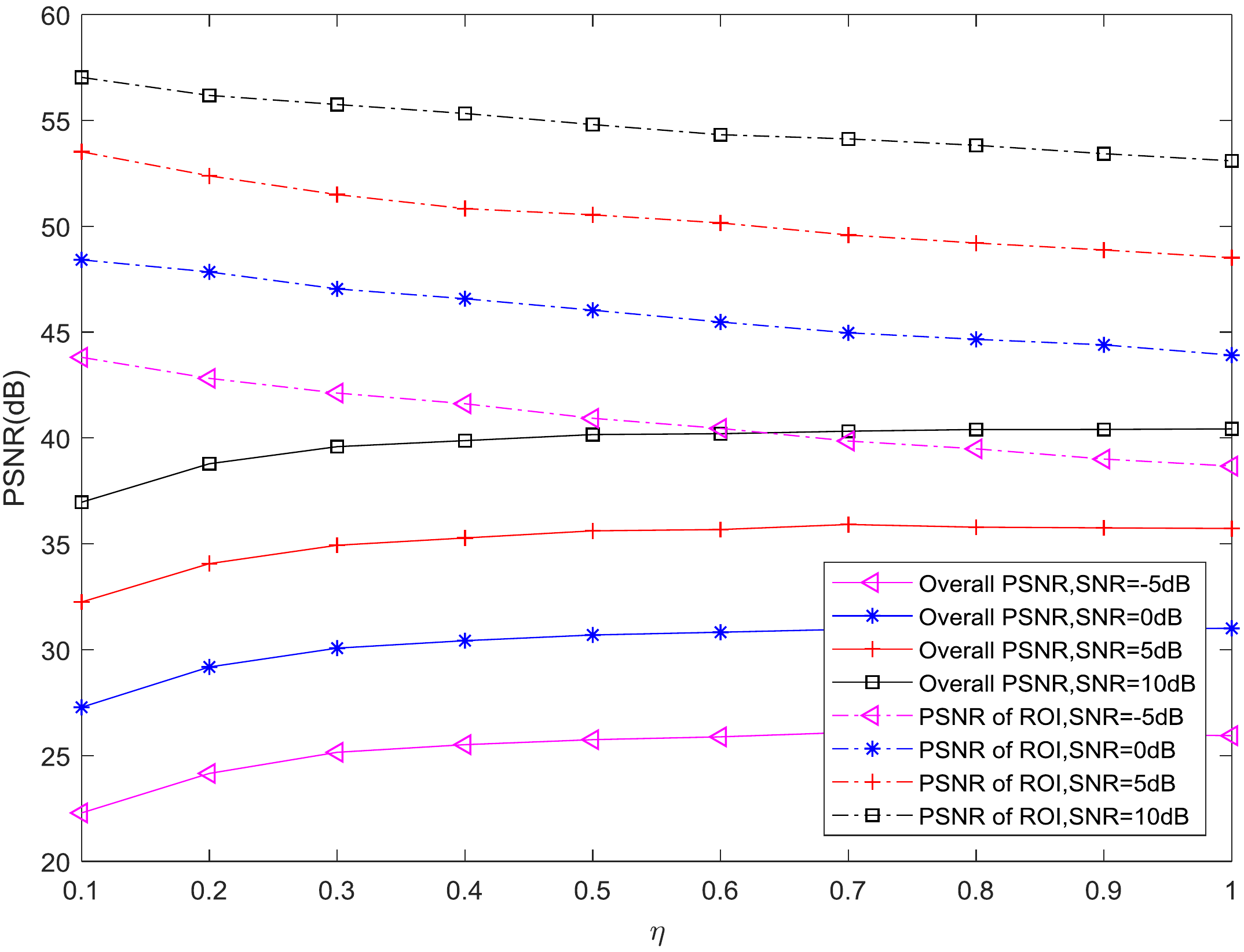}
\caption{The impact of the preference parameter $\eta$ on video recovery quality under different channel conditions.}
\label{F6}
\end{figure}

From Fig. 6, one can see that the overall PSNR improves with the increase of $\eta$, while the PSNR of the ROI declines with the increase of the $\eta$ under different channel conditions. The larger $\eta$ is, the less transmission power the ROI will get. Hence, the protection of the ROI data will be weakened with the increase of $\eta$. Fig. 6 also shows that the essence of the unequal power allocation scheme is its capability of improving the quality of ROI region at the expense of the quality of non-ROI region.

\vspace{-2mm}
\subsection{Performance Analysis of Processing Time of Each Step}
We then perform a delay analysis to evaluate the processing time for the three main steps, i.e., 1) \emph{ROI extraction}, 2) \emph{side information compression}, and 3) \emph{unequal power allocation}. We evaluate the processing time of ROI extraction using three different types of GPUs including GTX 1070 [44], GTX 1080 Ti [45], and RTX 2080 Ti [46] (where GTX 1070 $<$ GTX 1080 Ti $<$ RTX 2080 Ti in terms of computation performance), on Ubuntu 16.04 system with 4-core CPU, 16G memory, Cuda 10.0, and Cudnn 7.0.5. The processing time of side information compression and unequal power allocation are evaluated using Matlab 2018b. The testing results are shown in Table IV.

\begin{table}[htbp!]
\centering
\caption{Processing time of each step.}
\scalebox{0.82}{
\begin{tabular}{|c|c|c|c|c|c|}
\hline
\multirow{3}{*}{Frame} & \multicolumn{5}{c|}{Time (s)} \\ \cline{2-6}
& \multicolumn{3}{c|}{ROI extraction} & \multicolumn{1}{c|}{Side information} & \multicolumn{1}{c|}{Unequal power} \\ \cline{2-4}
&1070 &1080Ti &2080Ti &\multicolumn{1}{c|}{compression} & \multicolumn{1}{c|}{allocation}  \\
\hline
Coastguard\#2 &0.02904 &0.01720 &0.01337 &0.00313 &0.01678 \\
\hline
Coastguard\#180 &0.02911 &0.01731 &0.01337 &0.00327&0.01645 \\
\hline
Highway\#2 &0.02910 &0.01718 &0.01336 &0.00304 &0.01562 \\
\hline
Highway\#180 &0.02910 &0.01729 &0.01338 &0.00316 &0.01552 \\
\hline
Container\#2 &0.02905 &0.01728 &0.01336 &0.00320 &0.01636 \\
\hline
Container\#180 &0.02894 &0.01720 &0.01335 &0.00325 &0.01628 \\
\hline
Foreman\#2 &0.02897 &0.01721 &0.01338 &0.00301 &0.01565 \\
\hline
Foreman\#180 &0.02908 &0.01720 &0.01339 &0.00342 &0.01561 \\
\hline
Akiyo\#2 &0.02912 &0.01724 &0.01339 &0.00306 &0.01529 \\
\hline
Akiyo\#180 &0.02919 &0.01723 &0.01340 &0.00319&0.01565 \\
\hline
Hall\#2 &0.02911 &0.01715 &0.01333 &0.00312 &0.01569 \\
\hline
Hall\#180 &0.02911 &0.01718 &0.01340 &0.00315 &0.01578 \\
\hline
{\color{blue}Average} &{\color{blue}0.02908} &{\color{blue}0.01722} &{\color{blue}0.01337} &{\color{blue}0.00317} &{\color{blue}0.01589} \\
\hline
\end{tabular}}
\label{T2}
\end{table}

From Table IV, we can see that the processing time of \emph{ROI extraction} is mainly determined by the GPU performance, regardless of the image contents. The average processing time of each image decreases with the improvement of the GPU performance. With the decrease of GPU price, high-performance GPUs are becoming more and more widely used in video transmission to address the issues such as image recognition, image enhancement, etc.

The average processing time of \emph{side information compression} and \emph{unequal power allocation} are 0.00317s and 0.01589s, respectively, which indicates good performance compared to the existing digital transmission schemes. With GPU of RTX 2080 Ti, the average total processing time of each frame is 0.03243s, which means that the proposed scheme can support a frame rate of 30 fps.
\vspace{-2mm}
\subsection{Performance Analysis of Reconstructed Images}
In this part, we compare the performance of the proposed ROIC-Cast with three benchmark methods including SoftCast, DAC-RAN, and KMV-Cast in terms of the PSNR and the visual quality of the reconstructed images. SoftCast is the first pseudo analog scheme in which the optimal transmission power allocated for each block is determined by the variance. DAC-RAN can utilize the correlation information retrieved from the video big data as the prior knowledge to assist with the video demodulation. KMV-Cast exploits the hierarchical Bayesian model to further eliminate the mutual interference in DAC-RAN.

We focus on two cases: One has a strong correlation between the transmitted frame (i.e., \#2) and the reference frame(i.e., \#1), and the other has weak correlation (between \#180 and \#1). Since the insights contained in the reconstructed results of the six testing video sequences are similar, we select ``Coastguard'' as an example to explain its performance here. The image reconstruction results are shown in Fig. 7. For ease of comparison, the PSNR values of all reconstructed images are shown in Table V where the best performance under each simulation condition has been highlighted.

There are 5 rows in Fig. 7, of which the first row contains three baselines chosen from ``Coastguard'', including: 1) the reference frame found in the cloud database; 2) the second frame to be transmitted, and 3) the $180^{th}$ frame to be transmitted, respectively. The three baselines are used for the comparisons of image reconstruction results under different conditions which are shown in the remaining four rows of Fig. 7.

\begin{figure*}[htbp!]
\centering
\includegraphics[width=1\textwidth]{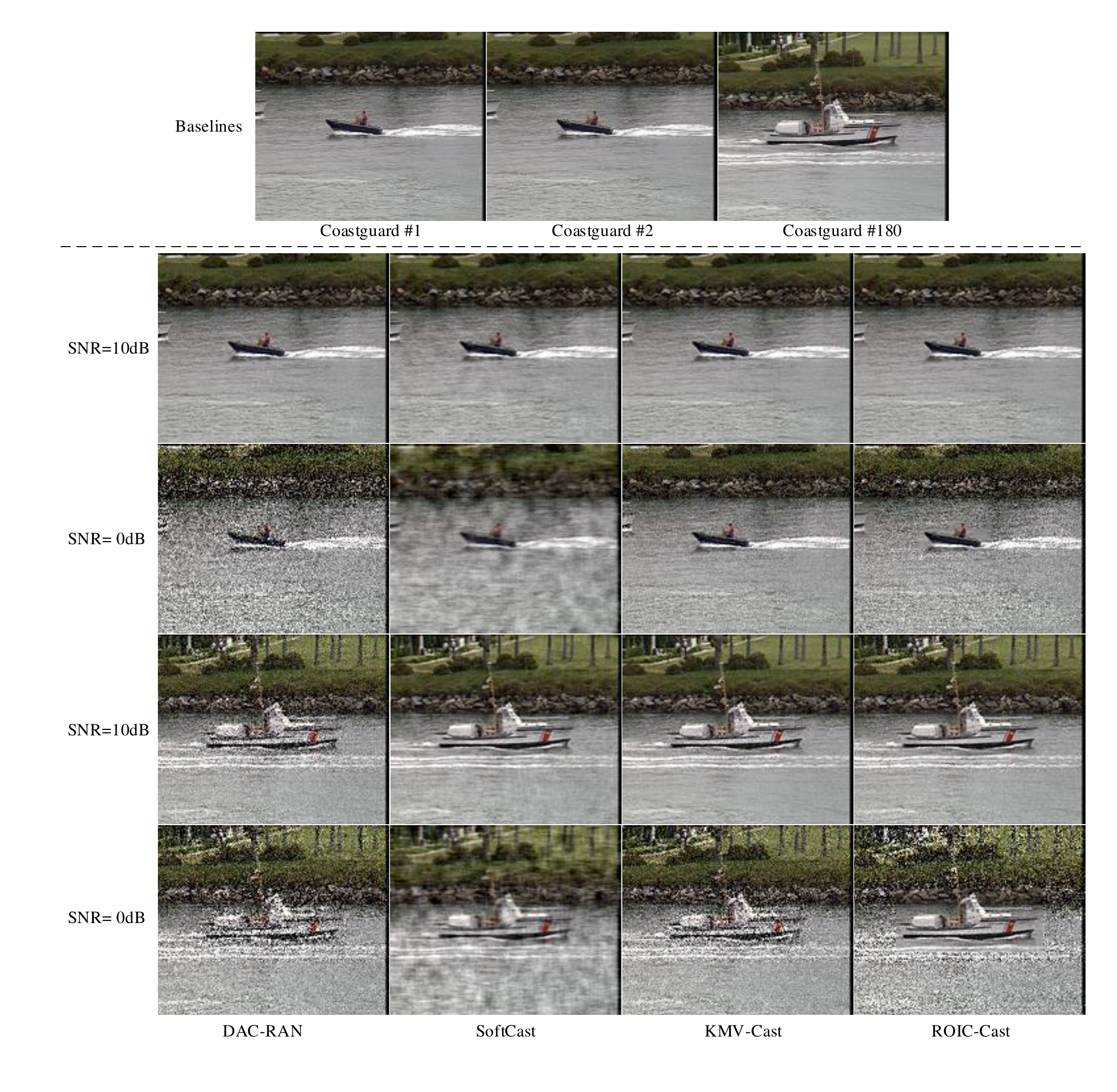}
\caption{Row 1: three baselines chosen from Coastguard for comparisons of image reconstruction results; Rows 2-5: the reconstructed images of different methods under different conditions.}
\label{F7}
\end{figure*}

\begin{table*}[htbp!]
\centering
\caption{PSNR performance of different methods under different conditions.}
\scalebox{0.92}{
\begin{tabular}{|c|c|c|c|c|c|c|c|c|c|}
\hline
\multirow{4}{*}{Dataset}& \multirow{4}{*}{Methods} & \multicolumn{4}{c|}{Strong correlation case \#2} & \multicolumn{4}{c|}{Weak correlation case \#180} \\ \cline{3-10}
& & \multicolumn{2}{c|}{SNR=10dB} &\multicolumn{2}{c|}{SNR=0dB} &\multicolumn{2}{c|}{SNR=10dB} &\multicolumn{2}{c|}{SNR=0dB} \\ \cline{3-10}
& & overall PSNR & PSNR of ROI & overall PSNR & PSNR of ROI & overall PSNR & PSNR of ROI & overall PSNR & PSNR of ROI \\
& & (dB) & (dB) & (dB) & (dB) & (dB) & (dB) & (dB) & (dB) \\
\hline
\multirow{4}{*}{Coastguard} & DAC-RAN &24.83&37.26&14.25&26.68&24.28&31.11&14.27&21.09 \\ \cline{2-10}
&SoftCast&33.70&45.10&24.91&36.07&31.64&39.23&{\color{blue}23.40}&29.51 \\ \cline{2-10}
&KMV-Cast&{\color{blue}40.39}&53.14&{\color{blue}30.87}&43.70&{\color{blue}32.54}&36.16&22.62&26.13 \\ \cline{2-10}
&ROIC-Cast&38.84&{\color{blue}57.24}&29.17&{\color{blue}48.64}&32.24&{\color{blue}40.26}&22.41&{\color{blue}30.38} \\
\hline
\multirow{4}{*}{Highway} & DAC-RAN &19.99&30.46&9.57&22.20&20.06&32.60&9.54&23.39 \\ \cline{2-10}
&SoftCast&33.28&45.81&24.29&35.95&33.29&46.41&24.43&37.97 \\ \cline{2-10}
&KMV-Cast&{\color{blue}37.44}&45.60&{\color{blue}28.54}&35.84&{\color{blue}37.42}&45.15&{\color{blue}28.48}&35.55 \\ \cline{2-10}
&ROIC-Cast&36.98&{\color{blue}51.31}&28.01&{\color{blue}43.45}&37.22&{\color{blue}52.22}&28.21&{\color{blue}44.40} \\
\hline
\multirow{4}{*}{Container} & DAC-RAN &24.86&28.39&13.24&17.28&24.83&27.96&13.55&16.50 \\ \cline{2-10}
&SoftCast&31.49&37.00&23.39&27.76&31.67&37.19&23.26&28.04 \\ \cline{2-10}
&KMV-Cast&{\color{blue}44.19}&47.59&{\color{blue}38.11}&40.69&{\color{blue}36.67}&37.71&{\color{blue}27.00}&27.95 \\ \cline{2-10}
&ROIC-Cast&43.39&{\color{blue}48.84}&36.24&{\color{blue}43.80}&36.24&{\color{blue}41.84}&26.65&{\color{blue}32.15} \\
\hline
\multirow{4}{*}{Foreman} & DAC-RAN &23.84&26.43&12.51&14.61&23.40&24.28&12.90&13.72 \\ \cline{2-10}
&SoftCast&32.23&34.56&23.89&26.46&32.88&34.30&24.09&25.64 \\ \cline{2-10}
&KMV-Cast&{\color{blue}39.18}&40.07&{\color{blue}30.38}&31.69&{\color{blue}34.75}&37.04&{\color{blue}25.06}&27.56 \\ \cline{2-10}
&ROIC-Cast&36.87&{\color{blue}41.99}&27.45&{\color{blue}32.29}&29.39&{\color{blue}37.86}&19.59&{\color{blue}28.36} \\
\hline
\multirow{4}{*}{Akiyo} & DAC-RAN &24.40&25.06&13.92&14.57&24.48&24.95&13.91&14.34 \\ \cline{2-10}
&SoftCast&33.66&33.96&25.40&26.46&33.91&34.76&25.28&26.00 \\ \cline{2-10}
&KMV-Cast&{\color{blue}44.65}&45.47&{\color{blue}41.22}&41.88&{\color{blue}39.19}&39.35&{\color{blue}30.15}&30.20 \\ \cline{2-10}
&ROIC-Cast&44.50&{\color{blue}45.53}&40.67&{\color{blue}42.27}&38.37&{\color{blue}39.60}&29.53&{\color{blue}30.66} \\
\hline
\multirow{4}{*}{Hall} & DAC-RAN &24.36&32.92&13.85&22.17&24.51&33.86&13.98&23.08 \\ \cline{2-10}
&SoftCast&32.01&42.73&23.62&34.02&31.74&42.20&23.56&34.67 \\ \cline{2-10}
&KMV-Cast&{\color{blue}41.19}&44.11&{\color{blue}32.63}&35.87&{\color{blue}41.15}&44.92&{\color{blue}32.84}&35.46 \\ \cline{2-10}
&ROIC-Cast&39.42&{\color{blue}48.62}&30.70&{\color{blue}39.72}&39.79&{\color{blue}49.12}&31.03&{\color{blue}40.54} \\
\hline
\end{tabular}}
\label{T4}
\end{table*}

The second row and the third row of Fig. 7 show the simulation results under the condition that the transmitted frame has a strong correlation with the reference frame. One can see that the proposed scheme outperforms others. Compared with others, there is a gain of more than 4.1dB at high channel SNR (i.e., 10dB) and a gain of more than 4.9dB at low channel SNR (i.e., 0dB), in terms of the PSNR of ROI. From the second row, one can find that the four schemes all perform well in terms of the subjective visual quality at high SNRs. But in the case of low SNR, the ROI image part of the proposed ROIC-Cast is much clearer than other three schemes.

The fourth row and the fifth row of Fig. 7 show the weak correlation case. The proposed ROIC-Cast achieves more than 1.0dB of gain at high SNR (i.e., 10dB) and more than 0.9dB of gain at low channel SNR in terms of the PSNR of ROI, compared with other three schemes. We can also see from the fifth row that DAC-RAN performs the worst in terms of the subjective visual quality at high SNR (i.e., 10dB), while there is not much difference among ROIC-Cast, KMV-Cast, and SoftCast. From the fifth row, one can clearly see the ship belonging to the ROI part in the proposed scheme, instead of an obscure image in other schemes under low SNR.

We then make a more intuitive performance comparison for different frames (\#2, \#10, and \#180) of the testing sequences under different channel conditions. The results are shown in Table V. Specifically, we present the performance comparison result using Coastguard in Fig. 8. Because the proposed ROIC-Cast can allocate transmission power for ROI data and non-ROI data unequally, one can see from Table V and Fig. 8 that the proposed scheme can achieve the best performance in terms of PSNR of ROI and subjective visual quality, even though the overall PSNR may not be as good as KMV-Cast and SoftCast under different channel conditions. However, as the similarity between the transmitted frame and the reference frame decreases, the superiority of the ROIC-Cast over other three schemes also declines.
\begin{figure*}[htbp!]
\centering
\includegraphics[width=1\textwidth]{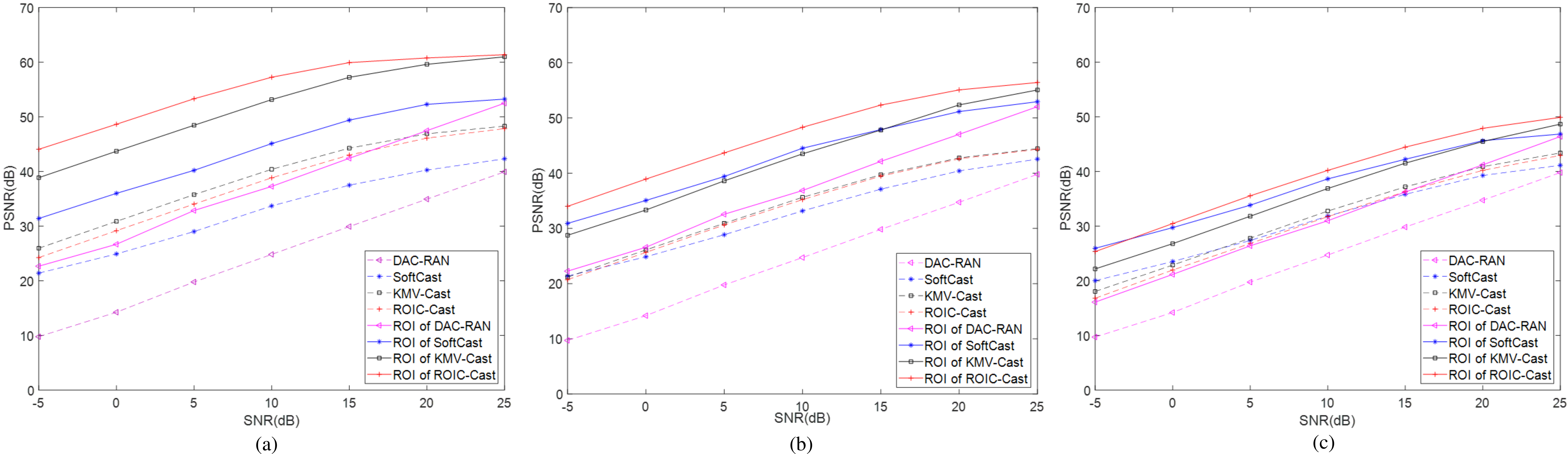}
\caption{The performance of different schemes under different SNRs (a) Coastguard (\#2); (b) Coastguard (\#10); (c) Coastguard (\#180).}
\label{F13}
\end{figure*}

\subsection{Performance analysis under the Rayleigh fading channel}
We also test and evaluate the performance of the proposed ROIC-Cast scheme under the channel conditions of Rayleigh fading. The simulation results are shown in Fig. 9 and Fig. 10.
\begin{figure}[htbp!]
\centering
\includegraphics[width=0.5\textwidth]{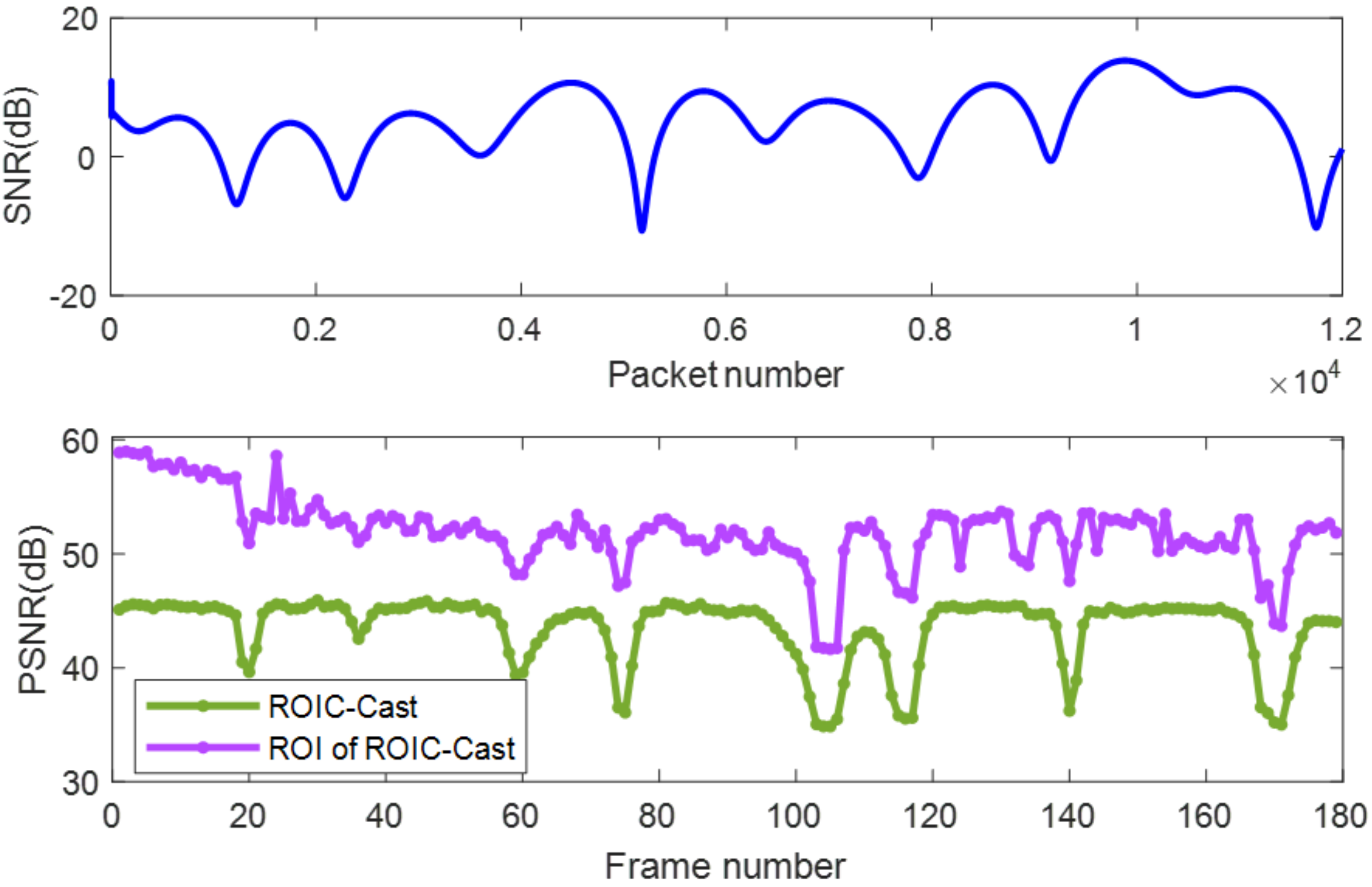}
\caption{Performance under Rayleigh fading channel. The top graph shows the SNR of the received packets and the bottom graph plots the overall PSNR of each frame and the PSNR of ROI part in each frame.}
\label{F1}
\end{figure}

the top graph in Fig. 9 shows the SNR of the received packets under the Rayleigh fading channel with an average of 5dB. The bottom graph of Fig. 9 plots the overall PSNR of each frame as well as the PSNR of ROI part in each frame. From Fig. 9, one can see that the proposed scheme can achieve adaptive PSNR performance in dramatically fluctuating channels. In order to reduce the effect of fading on the transmission quality, we apply {\emph{whitening}} (i.e., Hadamard matrix multiplication) to the transmitted signals. Whitening is similar to the conventional pseudo-random scrambling and interleaving of the bit-stream, but operates on the real input samples. The aim of whitening is to transform the input signal so that the transformed signal has the distribution of a memory-less Gaussian random variable. Therefore, whitening can mask fading and can improve performance beyond simple interleaving. From Fig. 9, we can also conclude that the proposed ROIC-Cast scheme can provide more protection for ROI part against non-ROI part. 

Fig. 10 shows the visual quality of four randomly selected reconstructed frames (\#$2$, \#$60$, \#$120$,\#$180$) of the testing sequence ``Coastguard''. From Fig. 9, we can find that the PSNR performance of the frame \#$60$ suffers from a sharp decline in terms of PSNR due to the channel fading. However, we can see from Fig. 10 that the frame \#$60$ can still provide a good visual quality. From Fig. 10, one can conclude that the proposed ROIC-Cast scheme can still achieve high-quality reconstructed frames although channel conditions vary a lot. Especially, we can clearly extract the ROI content in each frame. 
\begin{figure*}[htbp!]
\centering
\includegraphics[width=1\textwidth]{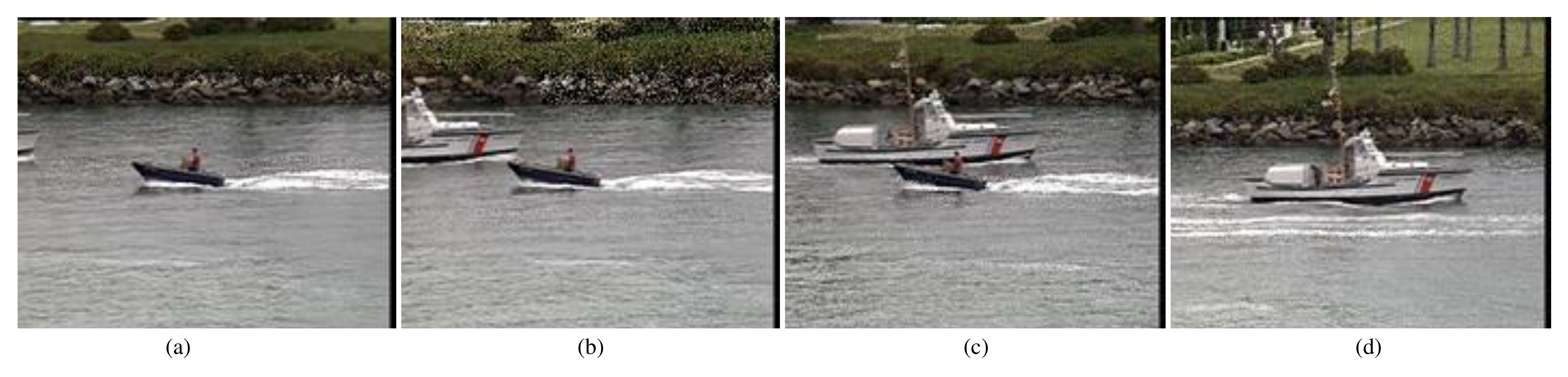}
\caption{The visual quality of four randomly selected reconstructed frames (\#$2$, \#$60$, \#$120$,\#$180$) of the testing sequence ``Coastguard'' in Rayleigh fading channel.}
\label{F2}
\end{figure*}

\section{Conclusions}
In this paper, a novel pseudo analog video transmission strategy called ROIC-Cast has been proposed to enhance the communication quality of ROI parts. Firstly, YOLOv2 has been utilized to extract the ROI from each video frame, and the ROI's location information is compressed by run-length coding. The video frames can be successfully classified into ROI and non-ROI pixel blocks. Secondly, a compression scheme for side information has been designed to support effective transmissions. Finally, an unequal power allocation algorithm has been proposed to protect the ROI transmissions, and ROI blocks can obtain more transmission power than non-ROI blocks. The simulation results have shown that the proposed ROIC-Cast scheme has the best performance compared with other three typical schemes, i.e., KMV-Cast, SoftCast, and DAC-RAN.

\section{Future Work}
The difficulty of achieving low-delay video communication lies in the large amount of video data and low bandwidth utilization. To handle the large data amount, we will study novel image video compression algorithms to further reduce the amount of data to be transmitted [47]. To solve the problem of low bandwidth utilization, we will combine pseudo analog transmission technology with cognitive radio networks (CRN), a promising technology to explore the idle spectrum and greatly improve the bandwidth utilization [48]. 

Our future work will weaken the effect of fluctuating channel conditions on video transmission quality from the following two approaches. On one hand, we can improve the channel quality of pseudo analog transmission by virtue of UAV's high mobility and line-of-sight (LoS) channel [49]. On the other hand, the combination of pseudo analogy technology and intelligent reflecting surface (IRS) will greatly improve the impact of channel fading and shadow on video quality degradation [50].

In addition, although some prototypes of IEEE 802.11 series have been developed, these prototypes cannot be used to verify the correctness of the new proposed pseudo analog wireless video transmission algorithms directly due to the limited modulation modes they can support. In our future work, we plan to design a software defined radio (SDR)-based pseudo analog wireless video transceiver which is completely transparent and allows users to learn all the implementation details [51, 52].

\vspace{-2mm}
\section{Acknowledgement}
The authors would like to thank all reviewers for their efforts in reviewing this manuscript.

\begin{IEEEbiography}[{\includegraphics[width=1in,height=1.25in,clip]{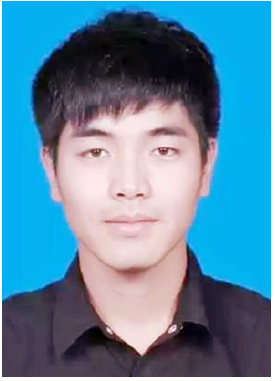}}]{Xiao-Wei Tang} (\emph{S'16, IEEE})
received the B.E. degree in Communication Engineering from Tongji University in 2016, where he is currently pursuing the Ph.D. degree. He has published several research papers on IEEE Transactions on Multimedia, IEEE Access, IEEE Globecom, and Mobile Networks \& Applications. He was a recipient of the Excellent Bachelor Thesis of Tongji University in 2016, the National Scholarship for Graduate Students by Ministry of Education of China in 2017, the Outstanding Students Award of Tongji University in 2017, the Outstanding Freshman Scholarship of Tongji University in 2018, the Chinese Government Scholarship by China Scholarship Council in 2019, the Outstanding Students Award of Tongji University in 2019, and the National Scholarship for Graduate Students by Ministry of Education of China in 2019. From Aug. 2019, he is doing research on UAV-enabled wireless video transmission in the Department of Electrical and Computer Engineering, the National University of Singapore, as a visiting scholar. His research interests include Pseudo-Analog Video Transmission, UAV Communication, Convex Optimization, and Deep Learning.
\end{IEEEbiography}

\begin{IEEEbiography}[{\includegraphics[width=1in,height=1.25in,clip]{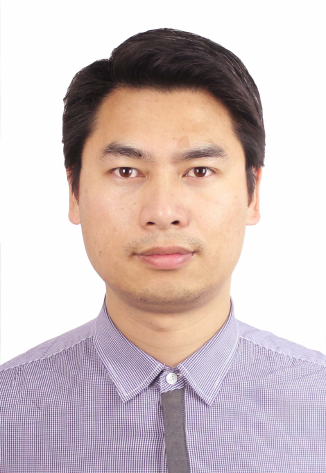}}]{Xin-Lin Huang}
(\emph{S'09-M'12-SM'16, IEEE}) is currently a professor and vice-head of the Department of Information and Communication Engineering, Tongji University, Shanghai, China. He received the M.E. and Ph.D. degrees in information and communication engineering from Harbin Institute of Technology (HIT) in 2008 and 2011, respectively. His research focuses on Cognitive Radio Networks, Multimedia Transmission, and Machine Learning. He published over 70 research papers and 8 patents in these fields. Dr. Huang was a recipient of Scholarship Award for Excellent Doctoral Student granted by Ministry of Education of China in 2010, Best PhD Dissertation Award from HIT in 2013, Shanghai High-level Overseas Talent Program in 2013, and Shanghai Rising-Star Program for Distinguished Young Scientists in 2019. From Aug. 2010 to Sept. 2011, he was supported by China Scholarship Council to do research in the Department of Electrical and Computer Engineering, University of Alabama (USA), as a visiting scholar. He was invited to serve as Session Chair for the IEEE ICC2014. He served as a Guest Editor for IEEE Wireless Communications and Chief Guest Editor for International Journal of MONET and WCMC. He serves as IG cochair for IEEE ComSoc MMTC, and Associate Editor for IEEE Access. He is a Fellow of the EAI.
\end{IEEEbiography}

\begin{IEEEbiography}[{\includegraphics[width=1in,height=1.25in,clip]{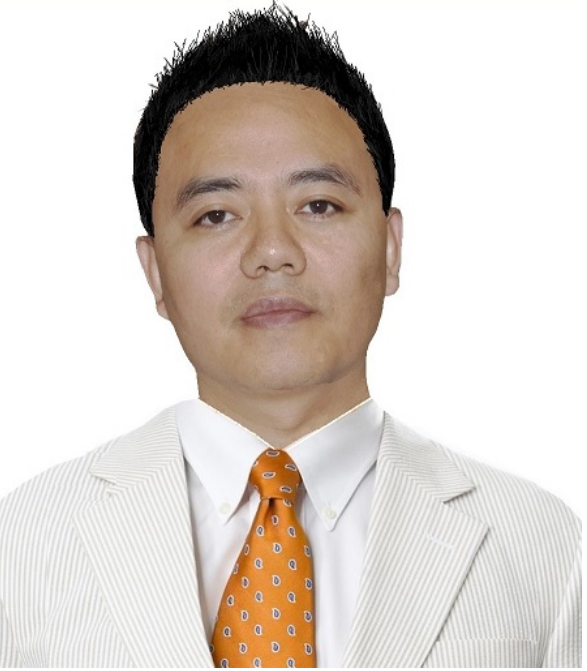}}]{Fei Hu}
(\emph{M'02, IEEE}) received the Ph.D. degree in signal processing from Tongji University, Shanghai, China, in 1999, and the Ph.D. degree in electrical and computer engineering from Clarkson University, Potsdam, NY, USA, in 2002. He is a Professor with the Department of Electrical and Computer Engineering, at the University of Alabama, Tuscaloosa, AL, USA. He has authored more than 200 journal/conference papers and book chapters in wireless networks, security, and machine learning. His research interests include cognitive radio networks, AI, and cyber security.
\end{IEEEbiography}

\begin{IEEEbiography}[{\includegraphics[width=1in,height=1.25in,clip]{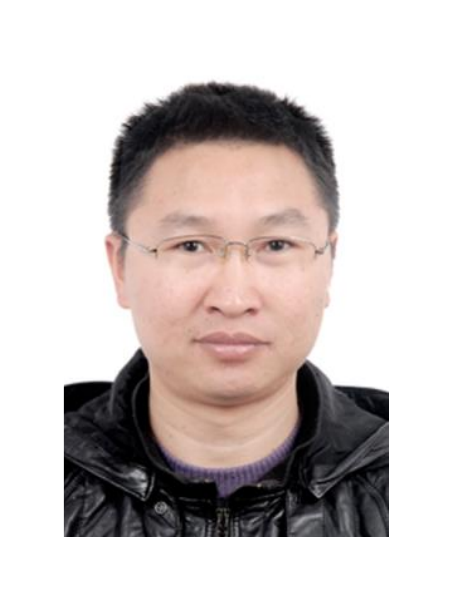}}]{Qingjiang Shi}
received the Ph.D. degree in communication engineering from Shanghai Jiao Tong University, Shanghai, China, in 2011. From September 2009 to September 2010, he visited Prof. Z.-Q. (Tom) Luo's research group, University of Minnesota, Twin Cities, MN, USA. He worked as a Research Scientist in 2011 at the Research and Innovation Center (Bell Labs China), Alcatel-Lucent Shanghai Bell Company, Ltd., China, and a Research Fellow in 2016 at Iowa State University, Ames, IA, USA. He became an Associate Professor in the School of Information and Science Technology, Zhejiang Sci-Tech University, Hangzhou, China, in 2013. Currently, he is a Professor in the School of Software Engineering, Tongji University, Shanghai, China. His current research interests include algorithm design and analysis for general areas of signal processing, wireless communications, and machine learning. He is currently an Associate Editor for the IEEE TRANSACTIONS ON SIGNAL PROCESSING, and was a TPC member for IEEE GLOBECOM/ICC and peer reviewers for a variety of the IEEE journals and conferences. He received the National Excellent Doctoral Dissertation Nomination Award in 2013, the Shanghai Excellent Doctoral Dissertation Award in 2012, and the Best Paper Award from IEEE PIMRC conference.
\end{IEEEbiography}

\vfill

\end{document}